\documentclass{appolb}%
\usepackage{amssymb,epsfig,amscd,amsmath,amsfonts}
\usepackage{amsmath}
\usepackage{amsfonts}
\usepackage{amssymb}
\usepackage{graphicx}%
\begin{document}

\title{Non exponential decays of hadrons \thanks{Presented at ``Excited QCD 2011,
20-25 February 2011- Les Houches (France)''}}
\author{Giuseppe Pagliara \address{Institut f\"{u}r Theoretische Physik,
Ruprecht-Karls-Universit\"at, Philosophenweg 16, D-69120, Heidelberg, Germany}
\and Francesco Giacosa
\address{Institute for Theoretical Physics, Johann Wolfgang Goethe
University, Max-von-Laue-Str.\ 1, D--60438 Frankfurt am Main, Germany} }
\maketitle

\begin{abstract}
We analyze the survival probability of unstable particles in the context of
quantum field theory. After introducing the spectral function of resonances,
we show that deviations from the exponential decay law occur at short times
after the creation of the unstable particle. For hadronic decays, these
deviations are sizable and could lead to observable effects.
\end{abstract}



\section{Introduction}

Deviations from the exponential decay law of unstable systems are a natural
consequence of the postulates of Quantum Mechanics
\cite{1978RPPh...41..587F,khalfin,2008JPhA...41W3001F}: for an unstable state,
whose average energy is finite, the survival probability for short times after
the \textquotedblleft creation\textquotedblright\ of the state is slower than
any exponential decay law. In other terms, if we introduce an effective, time
dependent decay rate $\gamma(t)=\frac{-1}{t}\mathrm{log}(p(t))$ one has that
for $t\rightarrow0^{+}$, $\gamma(0^{+})=0$ while at large times the standard
exponential decay law, $\gamma(t)\simeq\Gamma$, is obtained. The initial
temporal window for which deviations from the exponential law take place is
usually very small: it is of the order of $10^{-15}$ sec for electromagnetic
atomic decays \cite{Facchi:1999nq}. This explains why these deviations have
never been observed in experiments before 1997 \cite{raizen1} when, for the
first time, a cold atoms experiment has reported the evidence of such
deviations for \textquotedblleft bona fide\textquotedblright\ unstable states
(tunneling of atoms out of a trap). Previously, in \cite{1990PhRvA..41.2295I},
deviations from the exponential law have been reported within Rabi
oscillations. The short time deviations from the exponential law open up the
possibility of the so called Quantum Zeno effect \cite{1977JMP....18..756M,2008JPhA...41W3001F,Facchi:1999nq}:
by \textquotedblleft observing\textquotedblright\ the system with pulsed
measurements at short times after its preparation, the effective decay rate is
reduced and eventually it vanishes for continuous measurements (Quantum Zeno
paradox). Also this prediction has been recently confirmed within cold atoms
experiments \cite{raizen2} and, moreover, the so called Inverse Quantum Zeno
effect has also been observed: in this case the measuring apparatus leads to a
faster decay of the unstable state \cite{2000Natur.405..546K}.

A natural question concerns the existence of deviations from the exponential
law also in the context of Relativistic Quantum Field Theory (RQFT) which is
the right theoretical frame for describing unstable particles. In the
perturbative approaches presented in \cite{1993PhRvL..71.2687B} no (or very
much suppressed) short-time deviations from the exponential law, and thus no
quantum Zeno effect, were found within RQFT. Here and in
Ref.~\cite{Giacosa:2010br} we reconsider the issue of the survival probability
in RQFT also by analyzing some subtleties one faces when trying to define
unstable particles, such as the problem of \textquotedblleft preparation of
the system\textquotedblright\ and of the fields redefinition. We will not
consider here the case of the fundamental Lagrangian of the Standard model but
we limit the discussion to a toy model superrenormalizable Lagrangian. We
indeed find that deviations from the exponential law occur also in a genuine
RQFT context and we discuss possible implications for hadronic decays.

\section{A model Lagrangian}

The toy Lagrangian we use to investigate the survival probability of an
unstable scalar particle $S$ decaying into two scalars $\varphi$ is given by:
\begin{equation}
\mathcal{L}=\frac{1}{2}(\partial_{\mu}S)^{2}-\frac{1}{2}M_{0}^{2}S^{2}%
+\frac{1}{2}(\partial_{\mu}\varphi)^{2}-\frac{1}{2}m^{2}\varphi^{2}%
+gS\varphi^{2}.\label{lag}%
\end{equation}
The interaction term $\mathcal{L}_{int}=gS\varphi^{2}$ is responsible for the
decay process $S\rightarrow\varphi\varphi$, whose tree-level decay rate reads:
\begin{equation}
\Gamma_{S\varphi\varphi}^{\text{t-l}}=\frac{\sqrt{\frac{M_{0}^{2}}{4}-m^{2}}%
}{8\pi M_{0}^{2}}(\sqrt{2}g)^{2}\text{ .}\label{tl1}%
\end{equation}
The `naive', tree-level expression of the survival probability $p(t)$ for the
resonance $S$ created at $t=0$ is $p_{\text{t-l}}(t)=e^{-\Gamma_{S\varphi
\varphi}^{\text{t-l}}t}$ and the tree-level expression of the mean life time
is $\tau_{\text{t-l}}=1/\Gamma_{S\varphi\varphi}^{\text{t-l}}$. Here we
interpret our Lagrangian as an effective model to describe the decays of
hadrons; it is therefore quite natural to introduce a cutoff $\Lambda$ on the
energy of the particles of the typical mass scale of strongly interacting
particles i.e. $\Lambda\sim1$ GeV. To introduce the cutoff in a more
consistent way one has to insert a nonlocal interaction in the Lagrangian
\cite{Giacosa:2007bn}:%

\begin{equation}
\mathcal{L}=\frac{1}{2}(\partial_{\mu}S)^{2}-\frac{1}{2}M_{0}^{2}S^{2}%
+\frac{1}{2}(\partial_{\mu}\varphi)^{2}-\frac{1}{2}m^{2}\varphi^{2}%
+\mathcal{L}_{int}\text{ ,}%
\end{equation}%
\begin{equation}
\mathcal{L}_{int}=gS(x)\int\mathrm{d}^{4}\mathrm{y}\varphi(x+y/2)\varphi
(x-y/2)\Phi(y)\text{ ,}%
\end{equation}
where $\Phi$ is a form factor whose Fourier transform, $f_{\Lambda}(q)=\int
d^{4}y\Phi(y)e^{-iyq}$, appears in the loop integrals and regularizes the
divergences. (For nonlocal Lagrangians see also Refs. \cite{nl} and refs.
therein.) In this work we will consider the case of a sharp cutoff and the
case of a smooth form factor. An intermediate step to obtain the survival
probability is the computation of the self energy which reads:
\begin{equation}
\Sigma(x=\sqrt{p^{2}},m)=-i\int\frac{d^{4}q}{(2\pi)^{4}}\frac{f_{\Lambda
}(q^{0},\overrightarrow{q})^{2}}{\left[  (q+p/2)^{2}-m^{2}+i\varepsilon
\right]  \left[  (-q+p/2)^{2}-m^{2}+i\varepsilon\right]  }%
\end{equation}
and modifies the propagator $\Delta_{S}$ of the unstable particle as usual:
\begin{equation}
\Delta_{S}(p^{2})=\left[  p^{2}-M_{0}^{2}+(\sqrt{2}g)^{2}\Sigma(p^{2}%
)+i\varepsilon\text{ }\right]  ^{-1}\text{.}%
\end{equation}

\section{Spectral functions and survival probabilities}

Similarly to the standard derivation within Quantum Mechanics, also in RQFT,
the survival probability can be obtained by projecting the initial unstable
state onto the energy eigenstates. In turn, this corresponds to the
calculation of the spectral function $d_{S}(x)$ of the scalar field $S$ which
is proportional to the imaginary part of the propagator:%
\begin{equation}
d_{S}(x=\sqrt{p^{2}})=\frac{2x}{\pi}\left\vert \lim_{\varepsilon\rightarrow
0}\mathrm{Im}[\Delta_{S}(p^{2})]\right\vert \text{ .}%
\end{equation}
The quantity $d_{S}(x)dx$ represents the probability that, in its rest frame,
the state $S$ has a mass between $x$ and $x+dx.$ It is correctly normalized
for each $g$, $\int_{0}^{\infty}d_{S}(x)dx=1$ and reproduces the limit
$d_{S}(x)=\delta(x-M_{0})$ for $g\rightarrow0$
\cite{Achasov:2004uq,Giacosa:2007bn}. Notice that there are situations in
which the the spectral function can be directly pin down by data because the
the background is small and well understood: the decay $\phi\rightarrow
\gamma\pi^{0}\pi^{0}$ through the intermediate $a_{0}(980)$ and $f_{0}(980)$
mesons, the similar decay of the $j/\psi$ charmonium, or the hadronic decay of
the $\tau$ lepton into $\nu\pi\pi,$ dominated by the $\rho$ meson for an
invariant $\pi\pi$ mass close to $\rho$ mass (e.g. \cite{Giacosa:2008st}).

The probability amplitude $a(t)$ and the survival probability $p(t)$ can be
then expressed as
\begin{equation}
a(t)=\int_{-\infty}^{+\infty}\mathrm{dx}\,\,d_{S}(x)e^{-ixt}\text{ ,
}p(t)=\left\vert a(t)\right\vert ^{2}\text{ .} \label{p(t)}%
\end{equation}
The condition $p(0)=1$ is fulfilled in virtue of the normalization of
$d_{S}(x)$.

Let us now study the first derivative of $p(t)$. We obtain that $p^{\prime
}(t=0)=0$ as a consequence of the fact that the integral $\int_{0}^{\infty
}x\,d_{S}(x)dx$ is finite and real (it is the mean mass $\left\langle
M\right\rangle $, a reasonable definition for the mass of a resonance
\cite{Giacosa:2007bn}). This, in turn, implies that the function
$\gamma(t)=\frac{-1}{t}\ln p(t)$ vanishes for $t\rightarrow0^{+}$:%
\begin{equation}
\lim_{t\rightarrow0^{+}}\gamma(t)=-\lim_{t\rightarrow0^{+}}\frac{p^{\prime
}(t)}{p(t)}=0.
\end{equation}
We can therefore conclude that the quantum Zeno effect is perfectly possible
in the present RQFT context.

\begin{figure}[ptb]
\begin{centering}
\epsfig{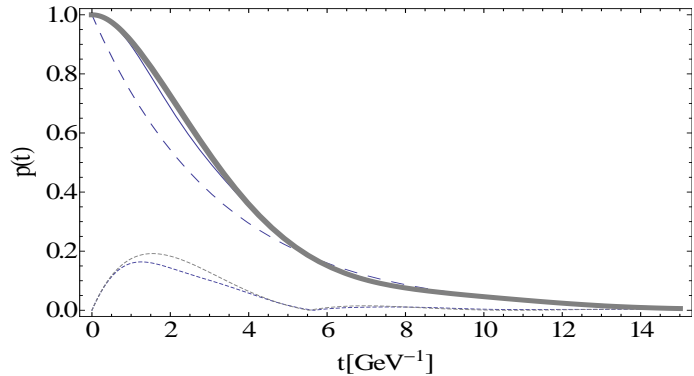}
\caption{Survival probability as a function of time. The solid line corresponds to the choice of a sharp cutoff, the thick gray line
to a smooth form factor, the dashed line to the exponential decay law and the dotted black and gray lines
are the differences between the survival probability as calculated from the spectral function and the exponential decay law. The deviations from the exponential law are quite sizable at short times.}
\end{centering}
\end{figure}

We show in Fig.~1 the survival probability for the case of a sharp cutoff
(solid line) a smooth form factor $f_{\Lambda}(q)=1/(1+(q/\Lambda)^{2})$
(thick gray line) and the standard exponential decay law (dashed line), here
$\Lambda=1.5$ GeV, $M_{0}=1$ GeV, $m=m_{\pi}$ and the tree level mean life
time $\tau_{t-l}=3.27$ GeV$^{-1}$ (this fixes $g$ in the two cases). Also
displayed are the differences between the survival probability as calculated
at one loop level and the tree level exponential decay law (dotted black and
gray lines). Notice that the time interval for which sizable deviations from
the exponential decay law occur is of the same order of magnitude of the mean
life time of the particle. This is an intriguing consequence of having
strongly interacting particles and could in principle lead to observable
effects, for instance in heavy ions collisions experiments. Moreover, the
difference between the sharp and smooth cutoff is very small: this fact
ensures that our results depend only slightly from the form of the cutoff function.

\section{Discussion and Conclusions}

There is an important issue that must be considered in connection with the
measurability of these deviations, which also correspond to the measurability
of the spectral function.

First, we notice that for very broad resonances, for which the deviation from
the exponential law are strong, one should also consider the mechanism by
which these resonances are created as, for instance, the scattering
$\varphi\varphi\rightarrow S\rightarrow\varphi\varphi$ \cite{Maiani:1997pd}.
One should introduce wave packets, with proper initial conditions, which
substantially overlap at $t=0$. In the framework of plane waves, the full
state of the system can be expressed in terms of the eigenstates of the
Hamiltonian $H_{0}$:
\begin{equation}
\left\vert s(t)\right\rangle =\sum_{\mathbf{k}}c_{\mathbf{k}}(t)\left\vert
\varphi_{\mathbf{k}}\varphi_{\mathbf{-k}}\right\rangle +c_{S}(t)\left\vert
S\right\rangle .\nonumber
\end{equation}
The coefficient $c_{S}(t)$ is vanishingly small for $t<<0$ and only for
$t\simeq0$ it becomes significant. If it were possible to tune the starting
conditions in such a way that $c_{S}(0)=1,$ we would have $\left\vert
s(t=0)\right\rangle =\left\vert S\right\rangle $ and the survival probability
of the resonance would be exactly the one presented in the previous section.
However, in general the state at $t=0$ is a superposition:
\begin{equation}
\left\vert s(0)\right\rangle =\sum_{\mathbf{k}}c_{\mathbf{k}}(0)\left\vert
\varphi_{\mathbf{k}}\varphi_{\mathbf{-k}}\right\rangle +c_{S}(0)\left\vert
S\right\rangle .\nonumber
\end{equation}
Further evolution implies:
\begin{align}
&  e^{-iHt}\left\vert s(0)\right\rangle =\nonumber\\
&  \sum_{\mathbf{k}}c_{\mathbf{k}}(0)e^{-iHt}\left\vert \varphi_{\mathbf{k}%
}\varphi_{\mathbf{-k}}\right\rangle +c_{S}(0)e^{-iHt}\left\vert S\right\rangle
=\nonumber\\
&  \sum_{\mathbf{k}}c_{\mathbf{k}}(0)e^{-iHt}\left\vert \varphi_{\mathbf{k}%
}\varphi_{\mathbf{-k}}\right\rangle +c_{S}(0)\left(  a(t)\left\vert
S\right\rangle +\left\vert \varphi\varphi\right\rangle \right)  .\nonumber
\end{align}
The amplitude $a(t)$ enters in a more general expression but it is not clear a
priori if the deviations from the exponential decay law are smeared out, in
the final \textquotedblleft measurement\textquotedblright\ of the decay
products, or if they could provide significant effects. A careful study would
be needed. Moreover, we plan also to investigate if the deviations from the
exponential decay law could indeed lead to observable effects also in Particle
Physics experiments.

The work of G.~P. is supported by the Deutsche Forschungsgemeinschaft (DFG)
under Grant No. PA 1780/2-1.


\end{document}